\documentclass[aps,prd,twocolumn,showpacs]{revtex4}

\usepackage{graphicx}
\usepackage{amsmath}
\usepackage{amssymb}

\usepackage{graphicx}
\usepackage{amsmath}
\usepackage{amssymb}
\usepackage{sistyle}
\usepackage{color}

\begin{document}

\title{Magnetic oscillations of the anomalous Hall conductivity}

\author{V.Yu.~Tsaran}
\affiliation{Department of Physics, Taras Shevchenko National Kiev University, 6 Academician Glushkov ave.,
Kiev 03680, Ukraine}

\author{S.G.~Sharapov}
\affiliation{Bogolyubov Institute for Theoretical Physics, National Academy of Science of Ukraine, 14-b
        Metrologicheskaya Street, Kiev 03680, Ukraine}

\date{\today }

\begin{abstract}
It is known that the Shubnikov--de Haas oscillations can  be observed in the Hall resistivity, although their amplitude
is much weaker than the amplitude of the diagonal resistivity oscillations.
Employing a model of two-dimensional massive Dirac fermions that exhibits anomalous Hall effect,
we demonstrate that the amplitude of the  Shubnikov--de Haas oscillations of the anomalous Hall conductivity is the same
as that of the diagonal conductivity.  We argue that the oscillations of the anomalous Hall conductivity
can be observed by studying the valley  Hall effect in graphene superlattices and the spin Hall effect in the low-buckled Dirac materials.
\end{abstract}









\maketitle

\section{Introduction}

It is almost 50 years since the Shubnikov--de Haas  (SdH) oscillations
were observed \cite{Fowler1966PRL} in a two-dimensional (2D) electron gas. In the next half century, the magnetoresistivity
measurements led to  remarkable discoveries  such as the quantum Hall effect and new Dirac materials such
as graphene. The diagonal part of the conductivity, $\sigma_{xx}$, exhibits oscillatory behavior (SdH effect)
as a function of the carrier concentration and/or magnetic field.  A theoretical description of the SdH oscillations is
usually provided by the formula \cite{Ando1974JPSJ,Ando1982RMP,Isihara1986JPC,Coleridge1989PRB,Gusynin2005PRB}
\begin{equation}
\label{sigma_xx}
\sigma_{xx}(B,\mu) = \frac{\sigma_0}{1 + \omega_c^2 \tau^2} \left[ 1 + \gamma f(\omega_c \tau) \frac{\tilde D (\mu)}{D_0(\mu)}
  \right],
\end{equation}
where
$\sigma_0$ is the conductivity in the absence of magnetic field $B$, $\omega_c = |eB|/m^\ast c$ is the cyclotron frequency with $m^\ast$
being the effective carrier mass and  $-e <0$ being the electron charge, and
$\tau$ is the relaxation time. In the last term in the brackets of Eq.~(\ref{sigma_xx}),
the numerical factor $\gamma =2$ for 2D electron gas \cite{Isihara1986JPC} and
$\gamma =1$ for the Dirac fermions \cite{Gusynin2005PRB},
$f(\omega_c \tau)$ is a smooth function of $\omega_c \tau$,
$\mu$ is the chemical potential,
$D_{0}(\mu)$ is the density of states (DOS) in the absence of magnetic field, and
$\tilde D (\mu)$ is the oscillatory part of the DOS that appears when an external magnetic field $B$
is applied perpendicular to the plane. The ratio $\tilde D (\mu)/D_0(\mu)$
at finite temperature can be written as follows:
\begin{equation}
\label{DOS-oscil}
\begin{split}
& \frac{\tilde D(\mu)}{D_0(\mu)} =  \\
& 2 \sum_{s =0}^{\infty} R_{T}(s \lambda) R_{D}(s)
\cos \left[2 \pi s \left(\frac{c S(\mu)}{2 \pi e \hbar B} + \frac{1}{2} + \beta \right)\right],
\end{split}
\end{equation}
where $S(\mu)$ is the electron orbit area in the momentum space, $\beta$ is the
topological part of the Berry phase,
\begin{equation}
\label{temp-factor}
R_T(s \lambda) = \frac{s \lambda}{\sinh s \lambda}, \qquad \lambda = \frac{2 \pi^2 k_B T}{\hbar \omega_c}
\end{equation}
is the temperature factor with $k_B$ being the Boltzmann constant, and
\begin{equation}
\label{Dingle-factor}
R_D(s) = \exp\left(-\frac{\pi s}{\omega_c \tau}  \right)
\end{equation}
is the Dingle factor. Equation~(\ref{DOS-oscil}) is written assuming that
the Landau levels have a Lorentzian shape with a width $\Gamma = \hbar/ (2 \tau)$ independent of energy or magnetic field.
The same assumption yields $f(\omega_c \tau) =1$ when the electrical conductivity (\ref{sigma_xx}) is calculated in the bare bubble approximation.
In general, $\Gamma$ and the real part of the electron self-energy have to be found by solving a self-consistent equation \cite{Ando1974JPSJ,Ando1982RMP,Isihara1986JPC}. When this is taken into account, the function
$f (\omega_c \tau)$ in Eq.~(\ref{sigma_xx}) acquires a nontrivial form. Nevertheless, the function $f (\omega_c \tau)$
turns out to be smooth and close to 1 for $\omega_c \tau \gtrsim 1$  \cite{Ando1974JPSJ,Ando1982RMP,Isihara1986JPC,Coleridge1989PRB}.

The case of a 2D electron gas with parabolic dispersion, $\epsilon = p^2/(2 m^\ast)$,
corresponds to $S(\mu) = 2 \pi m^\ast \mu$, and the trivial phase, $\beta = 0$.
The massive Dirac fermions with the dispersion, $\epsilon = \pm \sqrt{v_F^2 p^2 + \Delta^2}$, are characterized by
$S(\mu) = \pi (\mu^2 - \Delta^2)/v_F^2$, where $v_F$ is the Fermi velocity and $\Delta$ is the gap in the quasiparticle
spectrum, $m^\ast = |\mu|/v_F^2$ with $\omega_c = v_F^2 |e B|/(c|\mu|)$, and the nontrivial phase,
$\beta = 1/2$.  In the latter case, the oscillatory behavior is only possible for $|\mu| > |\Delta|$.

It is not as widely known that the SdH oscillations can also be observed in the Hall resistivity \cite{Coleridge1989PRB}.
The Hall resistivity is not just a monotonic or steplike function of $\mu$ and/or $B$,
towards the quantum Hall regime,
but also contains the {\em oscillatory part}. Using the St\v{r}eda formula \cite{Streda-formula},
it can be shown that \cite{Isihara1986JPC,Coleridge1989PRB}
the Hall conductivity reads
\begin{equation}
\label{Hall-normal-osc}
\begin{split}
\sigma_{i j}^{\textrm{H}}(B,\mu) = & - \frac{\epsilon_{ij}  \sigma_0 \omega_c \tau \, \mbox{sgn} \, (eB) \, \mbox{sgn} \, \mu}{1 +
\omega_c^2 \tau^2} \\
& \times
\left[ 1 - \frac{g(\omega_c \tau)}{\omega_c^2 \tau^2 } \frac{\tilde D (\mu)}{D_0(\mu)}
  \right].
\end{split}
\end{equation}
Here, $\epsilon_{ij}$ is the antisymmetric tensor ($i,j =x,y$ and $\epsilon_{12} =1$), and $g(\omega_c \tau)$
is a smooth function of $\omega_c \tau$.
It is difficult to measure the Hall resistivity oscillations not only because at low fields they are experimentally weaker than
the diagonal resistivity ones, but also since they must be separated from the large linear background. On the other hand,
at high fields the long-period oscillations with a rapidly varying amplitude have to be separated from  the monotonic background  \cite{Coleridge1989PRB}.

The {\em anomalous Hall effect} (AHE) is a phenomenon which consists of the
occurrence of the Hall effect in solids with broken time-reversal symmetry
in the absence of an external magnetic field.
When this field is applied, the Hall resistivity $\rho_{xy}$ is characterized by an empirical relation
\cite{Pugh1932PRev} (see Ref.~\cite{Nagaosa2010RMP} for a review),
\begin{equation}
\label{Hall-anom-empiric}
\rho_{xy} = R_0 B + R_{AH} M,
\end{equation}
where $R_0$ is the ordinary Hall constant, $R_{AH}$ is the anomalous Hall coefficient, and $M$ is the internal magnetization.
Equation~(\ref{Hall-anom-empiric}) applies to many materials over a broad range of external magnetic fields.

The discovery of Dirac materials such as graphene, silicene, etc. and topological insulators has considerably increased
the interest in AHE both from theory (see, e.g., Refs.~\cite{Sinitsyn2006PRL,Ezawa2012PRL,Ado2015EPL,Ando2015JPS}) and experiment
\cite{Mak2014Science,Gorbachev2014Science}.

Considering the relationship (\ref{Hall-anom-empiric}) from a perspective of the quantum magnetic oscillations,
it is reasonable to question whether these oscillations would emerge in the anomalous Hall conductivity.
In this article, we will show that in the Dirac materials  the oscillations of  the anomalous Hall conductivity can be as strong as
the SdH oscillations of the diagonal conductivity. To observe these oscillations in the existing Dirac materials, one should
measure either valley or spin Hall effects.

The paper is organized as follows. We begin by presenting
in Sec.~\ref{sec:model} the model describing  massive two-component Dirac fermions and
exhibiting the anomalous Hall effect. In
Sec.~\ref{sec:diagonal} we consider the SdH oscillations of the diagonal conductivity. It is shown that in addition to the
usual term, the diagonal conductivity also contains an anomalous term. The final expression for the SdH
oscillations of the anomalous Hall conductivity is presented in Sec.~\ref{sec:Hall}
(the calculational details and expressions for diagonal and normal Hall conductivity kernels
are given in Appendix~\ref{sec:Appendix-conductivity}, and the extraction of the oscillatory parts
is discussed in Appendix~\ref{sec:Appendix-oscillations}). In Sec.~\ref{sec:observation}, we suggest
the experiments that would allow one to observe the predicted oscillatory behavior.
In the conclusion given in Sec.~\ref{sec:conclusion}, the validity and applicability of the obtained results
is discussed in a  broader context.

\section{Model}
\label{sec:model}

We study the minimal model for AHE with broken time-reversal symmetry represented by
the two-component massive Dirac fermions in the presence of scalar Gaussian disorder. The
corresponding Hamiltonian density is
\begin{equation} \label{Dirac-Hamiltonian}
\mathcal{H}=   v_{F} \left[  \eta  \tau _{1} \left(\hat{p}_x + \frac{e}{c} A_x \right) +
\tau _{2}  \left(\hat{p}_y + \frac{e}{c} A_y \right)  \right] +\Delta \tau _{3}  -\mu \tau _{0},
\end{equation}%
where the Pauli matrices $\pmb{\tau} = (\tau_1, \tau_2,\tau_3)$ and  the unit matrix $\tau_0$
act in the pseudospin space, $\hat{p}_i =- i \hbar \partial_i$
with $i =x,y$ is the momentum operator,
and $\Delta/v_F^2$ is the Dirac mass.
The index $\eta = \pm$ distinguishes two inequivalent irreducible $2 \times 2$ representations of the Dirac algebra
in $2+1$ dimensions that correspond to the two independent valleys in the Brillouin zone
of the Dirac materials. The external magnetic field $\mathbf{B} = \mathbf{\nabla} \times \mathbf{A} = (0,0,B)$ is
applied perpendicular to the plane along the positive $z$ axis.

The main merit of the model (\ref{Dirac-Hamiltonian}) is that it not only
allows one to obtain simple analytical expressions, but also provides a deep insight into the underlying physics
\cite{Semenoff1984PRL,Haldane1988PRL}.
For example, for $B=0$ and $T=0$, the intrinsic (not induced by disorder) part of the Hall conductivity
reads \cite{Sinitsyn2006PRL} (see also Ref.~\onlinecite{Ando2015JPS} for a review)
\begin{equation}
\label{Hall-clean-final}
\sigma_{xy}^{\eta}=-\frac{e^{2} \mathrm{sgn}\, (\eta \Delta) }{4\pi \hbar }%
\begin{cases}
1, & |\mu |\leq |\Delta |, \\
|\Delta| /|\mu |, & |\mu |>|\Delta |.%
\end{cases}
\end{equation}%
The valley index $\eta$ in $\sigma_{xy}^{\eta}$ is omitted in what follows because it enters as a
product $\eta \Delta$ in all final expressions.
Yet a theoretical description of the AHE is a challenging problem. The presence of disorder makes the problem
rather complicated even in the absence of magnetic field. It has been recently shown \cite{Ado2015EPL}
that a widely used self-consistent Born approximation \cite{Sinitsyn2006PRL} fails when one
considers how the expression (\ref{Hall-clean-final}) is modified for $|\mu| > |\Delta|$ by  disorder.

Finishing the discussion of the model (\ref{Dirac-Hamiltonian}), we recall
that the AHE  may also be interpreted in terms of the motion of electrons in a fictitious magnetic field created
by the Berry curvature \cite{Tse2011PRB,Gorbachev2014Science} (see, also, Ref.~\onlinecite{Xiao2010RMP} for a review).

\section{Diagonal conductivity}
\label{sec:diagonal}

The presence of magnetic field significantly increases  the complexity of the problem. To obtain
analytical results, we model the smearing of Landau levels by Lorentzians with a constant width $\Gamma$
and use the Kubo formula with the bare vertex. We thus ignore the difference between the transport and single-particle
lifetimes, which is important for the analysis of the experimental data \cite{Coleridge1989PRB} unless
the scatterers are short range.

To illustrate the range of validity of this approximation, we
recapitulate the results for the diagonal conductivity of the Dirac fermions, $\sigma_{xx}$.
Using the Kubo formula, we show in Appendix~\ref{sec:Appendix-conductivity} that
the conductivity $\sigma_{xx}$ is the sum of the usual, $\sigma_{xx}^{\textrm{N}}$, and
the anomalous, $\sigma_{xx}^{\textrm{AH}}$, terms
\begin{equation}
\label{diagonal-conductivity}
\sigma_{xx}(B,\Delta,\mu) = \sigma_{xx}^{\textrm{N}}(B,\Delta,\mu) + \sigma_{xx}^{\textrm{AH}}(B,\Delta,\mu)
\end{equation}
with
\begin{equation}
\label{Kubo-diag}
\begin{split}
& \left\{
\begin{array}{c}
\sigma_{xx}^{\textrm{N}}(B,\Delta,\mu) \\
\sigma_{xx}^{\textrm{AH}}(B,\Delta,\mu) \\
\end{array}%
\right\}
=  \frac{e^{2}}{\hbar }
\left\{
\begin{array}{c}
1 \\
\mbox{sgn} \, (eB) \\
\end{array}%
\right\} \\
& \times
 \int \limits_{-\infty }^{\infty } d \epsilon
\left[ -\frac{\partial n_F (\epsilon)}{\partial \epsilon} \right]
\left\{
\begin{array}{c}
\mathcal{D}_{\textrm{N}}(\epsilon ,\mathcal{B},\Gamma,\Delta ) \\
\mathcal{D}_{\textrm{AH}}(\epsilon ,\mathcal{B},\Gamma,\Delta )  \\
\end{array}%
\right\}.
\end{split}
\end{equation}%
Here, $\mathcal{D}_{\textrm{N,AH}}$ is the kernel for the corresponding conductivity,
$n_F ( \epsilon ) = 1/[\exp((\epsilon-\mu)/k_B T)+1 ]$ is the Fermi distribution,
and we denoted $\mathcal{B} = \hbar v_F^2|eB|/c$.
The kernel $\mathcal{D}_{\textrm{N}}$ for the first term of Eq.~(\ref{diagonal-conductivity}), $\sigma_{xx}^{\textrm{N}}$,
was derived in Ref.~\onlinecite{Gorbar2002PRB} and we
include it for completeness in Appendix~\ref{sec:Appendix-conductivity} as  Eq.~(\ref{D-N}).
The SdH oscillations of this term were extracted in Ref.~\cite{Gusynin2005PRB}, so that  it can
be rewritten in the form (\ref{sigma_xx}) with $\gamma=1$, $f(\omega_c \tau)=1$, and $\sigma_0 = e^2 \tau (\mu^2 - \Delta^2)/(4 \pi \hbar^2 |\mu|)$
for $|\mu|> |\Delta|$. Using the relationship $|n^\ast| = (\mu^2 -\Delta^2)/(4 \pi \hbar^2 v_F^2)$ between
carrier imbalance $n^\ast$ and $\mu$ one can rewrite $\sigma_0$ in the standard form, $\sigma_0 = e^2 |n^\ast| \tau/m^\ast$.

The second term of Eq.~(\ref{diagonal-conductivity}), $\sigma_{xx}^{\textrm{AH}} \sim \eta \Delta B$,
is obtained in the present work. It is given by the bottom line of
Eq.~(\ref{Kubo-diag}) with the kernel (\ref{sigma_xx-anomalous-kernel})
from Appendix~\ref{sec:Appendix-conductivity}. It did not show up in the previous
studies \cite{Gorbar2002PRB,Gusynin2005PRB} because the summation over valleys, $\eta = \pm$, was done.

Expressions (\ref{sigma_xx}) and (\ref{Hall-normal-osc})
with the DOS, given by Eq.~(\ref{DOS-oscil}), the temperature, given by Eq.~(\ref{temp-factor}),
and the Dingle, given by Eq.~(\ref{Dingle-factor}), factors
describe both the nonrelativistic and Dirac fermions. It is worth noting that in the relativistic case
considered here, the effects related to the Landau level quantization can be observed at much higher temperatures
\cite{Novoselov2007Science}. This characteristic feature should be rather helpful for the experimental
observation of the oscillations discussed in what follows.

\section{Hall conductivity}
\label{sec:Hall}

Similarly to the diagonal conductivity,
we obtain in Appendix~\ref{sec:Appendix-conductivity}
that $\sigma_{xy}$ is the sum of the usual Hall, $\sigma_{xy}^{\textrm{H}}$, and
the anomalous Hall, $\sigma_{xy}^{\textrm{AH}}$, terms,
\begin{equation}
\label{Hall-general}
\sigma_{xy}(B,\Delta,\mu) = \sigma_{xy}^{\textrm{H}}(B,\Delta,\mu)+\sigma_{xy}^{\textrm{AH}}(B,\Delta,\mu)
\end{equation}
with
\begin{equation}
\label{Kubo}
\begin{split}
& \left\{
\begin{array}{c}
\sigma_{xy}^{\textrm{H}}(B,\Delta,\mu) \\
\sigma_{xy}^{\textrm{AH}}(B,\Delta,\mu) \\
\end{array}%
\right\}
=  \frac{e^{2}}{\hbar }
\left\{
\begin{array}{c}
\mbox{sgn} \, (eB) \\
1 \\
\end{array}%
\right\}
\\
& \times \int \limits_{-\infty }^{\infty } d \epsilon
\left[ -\frac{\partial n_F (\epsilon)}{\partial \epsilon} \right]
\left\{
\begin{array}{c}
\mathcal{A}_{\textrm{H}}(\epsilon ,\mathcal{B},\Gamma,\Delta ) \\
\mathcal{A}_{\textrm{AH}}(\epsilon ,\mathcal{B},\Gamma,\Delta )  \\
\end{array}%
\right\}.
\end{split}
\end{equation}%
Here $\mathcal{A}_{\textrm{H,AH}}$ is the kernel for the corresponding conductivity.
The normal Hall and anomalous Hall
terms obey the following symmetry relations
$\sigma_{xy}^{\textrm{H}}(B,\Delta,\mu) = -\sigma_{xy}^{\textrm{H}}(-B,\Delta,\mu) = \sigma_{xy}^{\textrm{H}}(B,-\Delta,\mu)=-\sigma_{xy}^{\textrm{H}}(B,\Delta,-\mu)$
and
$\sigma_{xy}^{\textrm{AH}}(B,\Delta,\mu) = \sigma_{xy}^{\textrm{AH}}(-B,\Delta,\mu) = -\sigma_{xy}^{\textrm{AH}}(B,-\Delta,\mu)=\sigma_{xy}^{\textrm{AH}}(B,\Delta,-\mu)$, respectively.
Notice that while the normal Hall conductivity $\sigma_{xy}^{\textrm{H}}(\mu)$ changes sign under the reversal
of carrier type, the anomalous Hall conductivity $\sigma_{xy}^{\textrm{AH}}(\mu)$ has the same
sign for both electrons and holes.
Furthermore, one can show that the Hall conductivity (\ref{Hall-general}) satisfies the
Onsager relation, $\sigma_{xy}(B,\Delta,\mu) = \sigma_{yx}(-B,-\Delta,\mu)$, that generalizes a usual
relation $\sigma_{xy}^{\textrm{H}} (B) = \sigma_{yx}^{\textrm{H}} (-B)$ for the normal Hall conductivity (\ref{Hall-normal-osc}).

The Hall kernel $\mathcal{A}_{\textrm{H}}$
was derived in Ref.~\onlinecite{Gusynin2006PRB} and we
include it for completeness in Appendix~\ref{sec:Appendix-conductivity}, Eq.~(\ref{AH}).
In the present work, we obtain the anomalous Hall kernel,
\begin{widetext}
\begin{equation} \label{kernel-AH}
\begin{split}
\mathcal{A}_{\textrm{AH}}(\epsilon, \mathcal{B},\Gamma,\Delta)= &\frac{\eta \Delta}{4 \pi^2}
\left\{
\frac{2\Gamma }{\mathcal{B}^2 + 4 \epsilon^2 \Gamma^2} \left[
-\mathcal{B} \frac{\mathcal{B} (\Delta^2 + \Gamma^2 -\epsilon^2) -
4 \Gamma^2 \epsilon^2}{[(\epsilon -\Delta)^2 + \Gamma^2] [(\epsilon +\Delta)^2 + \Gamma^2] }
+ 2 \epsilon \Gamma \, \mathrm{Im}  \psi\left(\frac{\Delta^2-(\epsilon+i\Gamma)^2}{2 \mathcal{B}}\right) \right]  \right. \\
& \left. - \frac{1}{|\Delta|} \left[ \arctan \frac{\epsilon + | \Delta|}{\Gamma} - \arctan \frac{\epsilon - | \Delta|}{\Gamma} \right]
\right\},
\end{split}
\end{equation}
\end{widetext}
where $\psi$ is the digamma function. Equation (\ref{kernel-AH}) includes all information about transitions between
Landau levels contributing to the conductivity. It allows a thorough analytic investigation of the various limiting cases.
First, using the asymptotic of the digamma function for the large arguments
given by Eq.~(\ref{psi-asymp}),
one can show that for low magnetic fields and $\Gamma \ll |\Delta -\epsilon|$,
the kernel can be approximated as follows:
\begin{equation}
\label{kernel-AH-appr}
\begin{split}
\mathcal{A}_{\textrm{AH}} (\mathcal{B}) \approx  -\frac{\eta \Delta}{4 \pi} \left[
\frac{\theta (\epsilon^2 - \Delta^2)}{|\epsilon| (1 + \mathcal{B}^2/(4 \epsilon^2 \Gamma^2))}
  + \frac{\theta (\Delta^2-\epsilon^2)}{|\Delta|}  \right].
\end{split}
\end{equation}
Accordingly, for $B=0$, Eq.~(\ref{kernel-AH-appr}) results in the anomalous Hall conductivity (\ref{Hall-clean-final}).
In general, the AHE shows robustness with respect to an external magnetic field when the chemical potential is inside the gap,
$|\mu| \leq |\Delta|$.   This resembles the results of Ref.~\cite{Zhang2014PRB},
where the robustness of the quantum spin Hall effect for the Bernevig-Hughes-Zhang model was studied.

The presence of $\Gamma$ as a factor in the first two terms of the kernel $\mathcal{A}_{\textrm{AH}}$ brings
the effects of the dissipation in $\sigma_{xy}$.
In the limit $\Gamma \to 0$, only the third term of the kernel (\ref{kernel-AH}) and, respectively, the last term of
Eq.~(\ref{kernel-AH-appr}) survive. This is in agreement with the results of Ref.~\onlinecite{Khalilov2015EPJC},
where the polarization operator in the $2+1$-dimensional quantum electrodynamics at finite $B$ and $\mu$
is derived. The corresponding term is associated in \cite{Khalilov2015EPJC} with the contribution of
virtual fermions, while the oscillatory term considered in the present work cannot be obtained in the clean limit.

The low-field expansion (\ref{kernel-AH-appr}) does not allow one to extract the quantum magnetic oscillations contained
in the digamma function  when the real part of its argument becomes negative.

The oscillations of $\mathcal{A}_{\textrm{AH}}$ in $1/B$ are extracted in
Appendix~\ref{sec:Appendix-oscillations}. Taking the low-temperature limit and integrating
over the energy [see Eq.~(\ref{integral})]
we arrive at the central result of this paper
\begin{equation}
\label{AH-oscil}
\begin{split}
\sigma_{ij}^{\textrm{AH}} (B,\Delta,\mu)&= - \frac{e^2 \epsilon_{ij} \mathrm{sgn}\, (\eta \Delta)}{4 \pi \hbar}
\Bigg\{  \theta(\Delta^2 - \mu^2)
\\
& + \frac{| \Delta|}{|\mu|} \frac{\theta(\mu^2-\Delta^2)}{1+\omega_c^2 \tau^2}
\left[ 1 + \gamma \frac{\tilde D(\mu)}{D(\mu)} \right] \Bigg\}.
\end{split}
\end{equation}

We find that the structure of Eq.~(\ref{AH-oscil}) resembles the diagonal conductivity (\ref{sigma_xx})
rather than the normal Hall term (\ref{Hall-normal-osc}).
One can see that the absence of the $\omega_c \tau$ prefactor in Eq.~(\ref{AH-oscil})
makes possible the observation of the oscillations of the anomalous Hall conductivity even in the low-field regime.
Moreover, the oscillating term is not damped by the $1/(\omega_c \tau)^2$
factor and has the same weight as the constant term for all strengths of the magnetic field.
It is shown in Appendix~\ref{sec:Appendix-oscillations}  that for $\Delta=0$ and small $\Gamma$, the weight of
the oscillatory part of the anomalous Hall conductivity, $g(\mu \tau)/(\omega_c \tau)^2 =|\mu|/(\omega_c^2 \tau \hbar)$,
which implies that for the normal Hall conductivity the amplitude of the oscillations is weakened as the relation
time $\tau$ increases. On the contrary, although the oscillatory term, $\sim \mbox{Im} \psi$, in the
initial expression (\ref{kernel-AH}) is proportional to $1/\tau^2$, in the final result (\ref{AH-oscil})
the relaxation time is present only in the $1/(1 + \omega_c^2 \tau^2)$ factor. One would expect
the anomalous Hall conductivity oscillations to come from the interplay
between the Berry curvature and the magnetic field.

\begin{figure}[h]
\centering{\includegraphics[width=8.5cm]{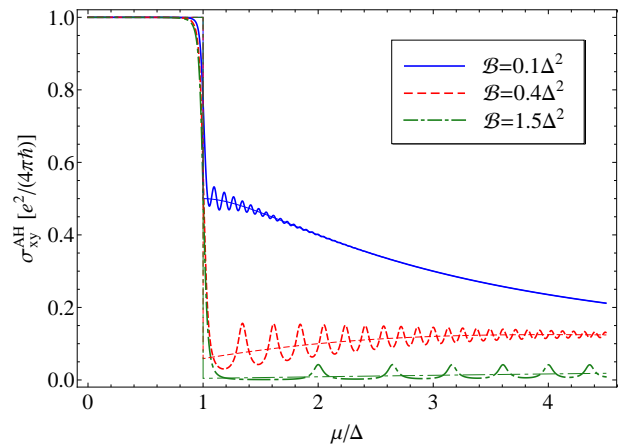}}
\caption{The anomalous Hall conductivity, $\sigma_{xy}^{\textrm{AH}} (\mu)$ in units of $e^2/(4 \pi \hbar)$ vs
chemical potential $\mu$ in units of $\Delta$ for temperature $T = 0$, scattering rate $\Gamma = 0.05 \Delta$ and $\eta=-1$.
$\mathcal{B} = 0.1 \Delta^2$ (solid blue line) , $\mathcal{B} = 0.4 \Delta^2$ (dashed red line), and
$\mathcal{B} = 1.5 \Delta^2$ (dash-dotted green line).
The thick and thin lines are plotted using Eqs.~(\ref{kernel-AH}) and (\ref{kernel-AH-appr}), respectively.}
\label{fig:1}
\end{figure}

\begin{figure}[h]
\centering{\includegraphics[width=8.5cm]{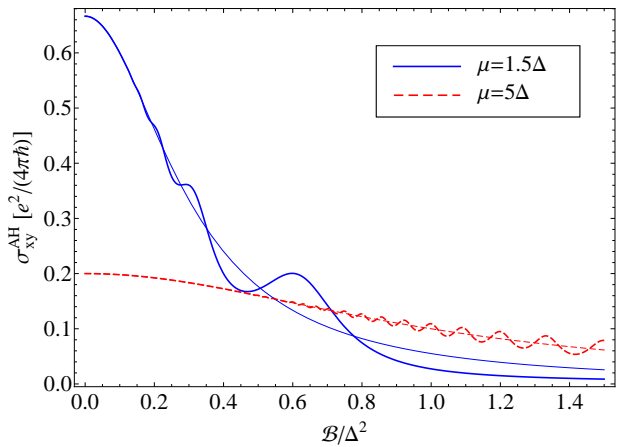}}
\caption{(Color online) The anomalous Hall conductivity, $\sigma_{xy}^{\textrm{AH}} (B)$ in units of $e^2/(4 \pi \hbar)$ vs
applied magnetic field $B$ in units of $\mathcal{B}/\Delta^2=\hbar v_F^2 |e B|/{c \Delta^2}$
for temperature $T = 0$, scattering rate $\Gamma = 0.05 \Delta$ and $\eta=-1$.
The solid blue line is for $\mu =1.5 \Delta$ and the dashed red one is for $\mu = 5 \Delta$.
The thick and thin lines are plotted using Eqs.~(\ref{kernel-AH}) and (\ref{kernel-AH-appr}), respectively.}
\label{fig:2}
\end{figure}

In Figs.~\ref{fig:1} and \ref{fig:2}, we show the results for
the  anomalous Hall conductivity $\sigma_{xy}^{\textrm{AH}} (B,\mu)$ as a function of
the chemical potential and magnetic field, respectively. The thick and thin lines are plotted
using the full  and approximated kernels given,
respectively, by Eqs.~(\ref{kernel-AH}) and (\ref{kernel-AH-appr}).
All curves were  obtained for the same impurity scattering $\Gamma =0.05 \Delta$ and the temperature $T = 0$.
The valley $\eta = -$ is chosen.
The curves in Fig.~\ref{fig:1}  differ in values of the magnetic field. Only the positive values of $\mu$
are shown because $\sigma_{xy}^{\textrm{AH}} (B,\mu)$ is an even function of $\mu$.
 Although the oscillatory behavior is present in all curves, it is seen at its best for an intermediate value of the
field, $\mathcal{B} = 0.4 \Delta^2$, shown by the dashed red curve. For lower value of the
field, $\mathcal{B} = 0.1 \Delta^2$, shown by the solid blue curve, the oscillations are quickly damped as the
chemical potential increases. The reason for this is that in the case of the Dirac fermions,
the Dingle factor $R_D$ (and, for a finite temperature, the temperature factor, $R_T$) depends on $|\mu |$.
For large fields, $\mathcal{B} = 1.5 \Delta^2$, shown by the dash-dotted green line, the field-dependent prefactor
in Eq.~(\ref{AH-oscil}) results in the smallness of $\sigma_{xy}^{\textrm{AH}}$. The system is expected
to enter the quantum Hall regime that cannot be described within the present approach.
The approximate kernel (\ref{kernel-AH-appr}) turns out to be in a good agrement with the exact one, reproducing
the nonoscillatory part of the Hall conductivity.
The curves in Fig.~\ref{fig:2} differ in values of the chemical potential $\mu$.
We find that the oscillatory behavior can be seen over a wide range of the magnetic fields. As discussed above,
it is better seen  when the values of $|\mu|$ are not much larger than the gap $\Delta$.
The approximate kernel (\ref{kernel-AH-appr}) correctly describes  the magnetic field dependence of the
nonoscillatory part of the Hall conductivity.

\section{Possibilities for experimental observation}
\label{sec:observation}

The model (\ref{Dirac-Hamiltonian}) represents a building
block of the full Hamiltonians of graphene and other Dirac materials. The corresponding electrical and spin
conductivities can be constructed by making the gap $\Delta$ dependent on the valley and spin indices $\eta$ and $\sigma$,
respectively, and summing over these degrees of freedom.

\subsection{Oscillations of the valley Hall conductivity in graphene}

The global $A/B$ sublattice asymmetry gap $2 \Delta \sim 350 \, \mbox{K}$ can be introduced in graphene
\cite{Hunt2013Science,Woods2014NatPhys,Chen2014NatCom,Gorbachev2014Science}
when it is placed on top of hexagonal boron nitride (G/hBN) and the crystallographic axes of graphene and hBN
are aligned.
For $B=0$, the normal Hall conductivity $\sigma_{xy}^{\textrm{H}}=0$ because of the time-reversal symmetry.
The anomalous parts of the conductivities, $\sigma_{xy}^{\textrm{AH}}$,  and  already mentioned, $\sigma_{xx}^{\textrm{AH}}$,
in the $\eta = +$ valley have the opposite sign to that in valley $\eta = -$ due to the same symmetry.
Although the full Hall conductivity $\sigma_{xy} =0$ in this case, the
nonlocal measurements \cite{Gorbachev2014Science} allow one to observe the charge neutral valley Hall effect.
The corresponding current, $j_{\textrm{v}} = j_{\eta = +} - j_{\eta = -} = \sigma_{xy}^{\textrm{v}} E $, where the valley Hall
conductivity  $\sigma_{xy}^{\textrm{v}} = 4 \sigma_{xy}^{\textrm{AH}}$, so that in the regime studied in \cite{Gorbachev2014Science},
$|\mu| < |\Delta|$, the value $\sigma_{xy}^{\textrm{v}} \approx 2 e^2/h$. It was also shown in \cite{Gorbachev2014Science}
that the nonlocal signal does not disappear in small magnetic fields.

It is important to stress that the value $\sigma_{xy}^{\textrm{v}}$ has to be extracted from the nonlocal measurements using the model
expression (see the  supplemental material of Ref.~\cite{Gorbachev2014Science})
$R_{NL}  \varpropto (\sigma_{xy}^{\textrm{v}})^2 \rho_{xx}^3,$
where $R_{NL}$ is the measured nonlocal resistance and $\rho_{xx}$ is the diagonal resistivity. The value $R_{NL}$
decays rapidly as the carrier concentration increases. Yet we hope that similar measurements can
be repeated for $|\mu| \gtrsim |\Delta|$ to observe the oscillations of the nonlocal resistivity and to extract from them the
oscillations related to the valley Hall conductivity.

The period of the oscillations, $B_F$, is given by
\begin{equation}
\label{period}
B_F[T] = \frac{(\mu^2 - \Delta^2) [\mbox{K}^2]}{1.77 *10^5 \mbox{K}^2 } = 0.15 (\mu^2/\Delta^2 -1)
\end{equation}
where we took $v_F = 10^6 \mbox{m/s}$.
The second equality in Eq.~(\ref{period}) is for $\Delta = 175 \mbox{K}$.
To obtain Eq.~(\ref{period}), we used that
the corresponding energy scale $\mathcal{B}$ expressed in the units of temperature reads
$\mathcal{B} \to \mathcal{B}/k_B^2 [\mbox{K}^2] = 8.85 \times 10^{-8} v_F^2[\mbox{m/s}] B[ \mbox{T}]$,
where $v_F$ and $B$ are given in m/s and Tesla, respectively.

We note that for a finite field, the normal Hall term $\sigma_{xy}^{\textrm{H}} \neq 0$, but
different symmetry properties of $\sigma_{xy}^{\textrm{AH}}(B,\mu)$ and $\sigma_{xy}^{\textrm{H}}(B,\mu)$
can be used to distinguish these contributions.

\subsection{Oscillations of the spin Hall conductivity in low-buckled Dirac materials}

The spin Hall conductivity of silicene and other low-buckled Dirac materials \cite{Liu2011PRL}
can be expressed  (see the first paper in Ref.~\onlinecite{Sinitsyn2006PRL}) in terms of the electric Hall conductivity $%
\sigma_{xy} (\Delta)$ for the two-component Dirac fermions by the relation
\begin{equation}  \label{Hall2spin-Hall}
\sigma_{xy}^{S_z} = - \frac{\hbar}{2e} \sum_{\eta, \sigma = \pm } \sigma
\sigma_{xy}^\eta (\Delta \to \Delta_{\eta \sigma}),
\end{equation}
with the valley- and spin-dependent gap
$\Delta _{\eta \sigma }=\Delta _{z}-\eta \sigma \Delta _{\text{SO}}$. Here, $\Delta _{\text{SO}}$ is the spin-orbit gap
caused by a strong intrinsic spin-orbit interaction in the low-buckled Dirac materials. It is a large value of  $\Delta _{\text{SO}}$, e.g.
$\Delta _{\text{SO}} \approx \SI{4.2}{meV}$ in silicene and $\Delta _{\text{SO}} \approx \SI{11.8}{meV}$ in germanene,
which makes the quantum spin Hall effect \cite{Kane2005PRL,Dyrdal2012PSS}
experimentally accessible in these materials. The gap $\Delta_z = E_z d$, where
$2 d$ is the separation between the two sublattices situated in different vertical planes,
can be tuned by applying the electric field $E_z$ perpendicular to the plane. This  results in
the on-site potential difference between the $A$ and $B$ sublattices resembling the case of graphene on hBN.

Adjusting the value of $E_z$, one can tune up  the case of the zero sublattice asymmetry gap, $\Delta_z =0$.
Then one finds that the contribution of the normal Hall term, $\sigma_{xy}^{\textrm{H}}$, cancels out and
only the anomalous Hall conductivity part, $\sigma_{xy}^{\textrm{AH}}$,
enters the resulting spin Hall conductivity,
\begin{equation}
\sigma_{xy}^{S_z}(\Delta_z=0) = \frac{2\hbar}{e}\sigma_{xy}^{\textrm{AH}}(\eta \Delta \to \Delta _{\text{SO}}).
\end{equation}
Here, for $B=0$ and $T=0$, the Hall conductivity $\sigma_{xy}^{\textrm{AH}}$ is given by Eq.~(\ref{Hall-clean-final}) and,
for finite $B$ and $|\mu|> |\Delta _{\text{SO}}|$, the oscillating Hall conductivity  is described by  Eq.~(\ref{AH-oscil}).
The period of the oscillations is given by Eq.~(\ref{period}).
The large value of  the spin-orbit gap in silicene and related materials makes it plausible to
observe not only the spin Hall effect, but also {\em oscillations} of the spin Hall conductivity.

\section{Conclusion}
\label{sec:conclusion}

Simultaneous measurements of the SdH oscillations in both the longitudinal and Hall resistivities
provide additional information on the transport phenomena. As shown in \cite{Coleridge1989PRB},
the proper usage of the transport and single particle lifetimes is necessary to  correctly analyze the low-field amplitudes
and phases of the SdH oscillations. The measurements at high fields showed the deviations that were attributed to localization.
This method, however, did not become widespread because of the difficulties of measuring the SdH oscillations of the Hall resistivity.

In the present paper, we demonstrated that in the case of the anomalous Hall conductivity, these difficulties are absent.
Definitely, the involved approximations   include neither the effects related to the difference between transport and
single-particle lifetimes nor the energy dependence of the transport relaxation time. If necessary, the former effects
can be included by using the transport time in the corresponding expressions for the conductivities, except to
the DOS part (\ref{DOS-oscil}), where the single-particle lifetime should be kept.
As shown recently in \cite{Ado2015EPL}, in the case of the anomalous Hall effect
in the Dirac materials, a consistent microscopical analysis of these effects seems to be even more complicated than in the case of
the 2D electron gas. Nevertheless, the obtained results clearly illustrate a principal possibility to observe
strong SdH oscillations of the anomalous Hall conductivity.

We also speculate that the SdH oscillations of the Hall conductivity may also be observed in topological insulators \cite{Ando2013JPSJ}.
Furthermore, the obtained results are applicable for doped anomalous Hall systems as well.
Although Eq.~(\ref{AH-oscil}) is not valid in the generic case, one may expect that the oscillations of the anomalous Hall
conductivity are described by the following expression:
$\sigma_{xy}^{\textrm{AH}} \sim 1/(1+ \omega_c^2 \tau^2) [1 + \gamma \tilde D(\mu)/D(\mu)]$.  The oscillations of the DOS
are still described by Eq.~(\ref{DOS-oscil}), but the electron orbit area $S(\mu)$ corresponds to the specific model
including the tight-binding case.

S.G.S. gratefully acknowledges E.V.~Gorbar, V.P.~Gusynin, V.M.~Loktev and A.A.~Varlamov for
helpful discussions.

\appendix

\section{Calculation of the electrical conductivity}
\label{sec:Appendix-conductivity}

The frequency-dependent electrical conductivity
tensor can be found using the Kubo formula
\begin{equation}
\label{conductivity-def}
\sigma_{ij}(\Omega)=\frac{\Pi_{ij}^R(\Omega+i0)-\Pi_{ij}^R(0)}{i\Omega},
\end{equation}
where $\Pi_{ij}^R(\Omega+ i0)$ is the retarded current-current correlation function.
The scheme of the calculation of  $\Pi_{ij}^R(\Omega+ i0)$ is described in detail in Appendix~A of
Ref.~\onlinecite{Gusynin2006PRB}. The only difference is that now the corresponding Dirac matrices are $2\times 2$ dimensional,
viz., $\gamma_\eta^\mu = (\tau_3, - i \eta \tau_2, i \tau_1)$ with $\mu=0,1,2$.
Thus we directly proceed to the following expression for the real part of the conductivity,
\begin{equation}
\label{sigma-full}
\sigma_{ij}(\Omega) = \sigma_{ij}^{\textrm{H}}(\Omega) + \sigma_{ij}^{\textrm{AH}}(\Omega),
\end{equation}
where the first term,
\begin{widetext}
\begin{equation} \label{sigma-Pi12}
\begin{split}
\sigma_{ij}^{\textrm{H}}(\Omega)&=\frac{e^2 \mathcal{B} }{4\pi^2 \,\Omega}\sum\limits_{n,m=0}^\infty
(-1)^{n+m+1}\int\limits_{-\infty}^\infty
d\epsilon\left\{\delta_{ij}(\delta_{n,m-1}+\delta_{n-1,m})
[n_F(\epsilon)-n_F(\epsilon^\prime)]{\rm
Re}[\Pi^1_{n,m}(\epsilon,\epsilon^\prime)-
\Pi^2_{n,m}(\epsilon,\epsilon^\prime)]\right.\\
&-\left. \epsilon_{ij}\,{\rm
sgn}(eB)(\delta_{n,m-1}-\delta_{n-1,m})[[n_F(\epsilon)-n_F(\epsilon^\prime)]
{\rm
Im}\Pi^1_{n,m}(\epsilon,\epsilon^\prime)-[n_F(\epsilon)+n_F(\epsilon^\prime)]
{\rm Im}\Pi^2_{n,m}(\epsilon,\epsilon^\prime)] \right\},
\end{split}
\end{equation}
was derived in \cite{Gusynin2006PRB,foot1}.
Here, $\epsilon^\prime =\epsilon + \Omega$ and
we set $\hbar = k_B=1$ in the Appendices.
A new term proportional to the $\eta \Delta$ is
\begin{equation} \label{sigma-Pi34}
\begin{split}
\sigma_{ij}^{\textrm{AH}}(\Omega)&=\frac{e^2 \mathcal{B} }{4\pi^2 \,\Omega}\sum\limits_{n,m=0}^\infty
(-1)^{n+m+1} \\
& \times \int\limits_{-\infty}^\infty
d\epsilon \left\{\delta_{ij} {\rm sgn}(eB) (\delta_{n,m-1}-\delta_{n-1,m})
[n_F(\epsilon)-n_F(\epsilon^\prime)]{\rm
Re}[\Pi^3_{n,m}(\epsilon,\epsilon^\prime)-
\Pi^4_{n,m}(\epsilon,\epsilon^\prime)]\right.\\
&-\left. \epsilon_{ij}(\delta_{n,m-1}+\delta_{n-1,m})[[n_F(\epsilon)-n_F(\epsilon^\prime)]
{\rm
Im}\Pi^3_{n,m}(\epsilon,\epsilon^\prime)-[n_F(\omega)+n_F(\omega^\prime)]
{\rm Im}\Pi^4_{n,m}(\epsilon,\epsilon^\prime)] \right\}.
\end{split}
\end{equation}
\end{widetext}
We introduced  in Eqs.~(\ref{sigma-Pi12}) and (\ref{sigma-Pi34}) the following functions
\begin{equation}
\begin{split}
\Pi^1_{n,m}(\epsilon,\epsilon^\prime)=\frac{(\epsilon^\prime+i\Gamma^\prime)(\epsilon-i\Gamma)
-\Delta^2}{[(\epsilon^\prime+i\Gamma^\prime)^2-M_n^2][(\epsilon-i\Gamma)^2-M_m^2]},\\
\Pi^2_{n,m}(\epsilon,\epsilon^\prime)=\frac{(\epsilon^\prime+i\Gamma^\prime)(\epsilon+i\Gamma)
-\Delta^2}{[(\epsilon^\prime+i\Gamma^\prime)^2
-M_n^2][(\epsilon+i\Gamma)^2-M_m^2]}
\end{split}
\end{equation}
and
\begin{equation}
\begin{split}
& \Pi^3_{n,m}(\epsilon,\epsilon^\prime)=\frac{\eta \Delta
[(\epsilon^\prime+i\Gamma^\prime) - (\epsilon-i\Gamma)]
}{[(\epsilon^\prime+i\Gamma^\prime)^2-M_n^2][(\epsilon-i\Gamma)^2-M_m^2]},\\
& \Pi^4_{n,m}(\epsilon,\epsilon^\prime)=\frac{
 \eta \Delta [(\epsilon^\prime+i\Gamma^\prime) - (\epsilon+i\Gamma)]
}{[(\epsilon^\prime+i\Gamma^\prime)^2
-M_n^2][(\epsilon+i\Gamma)^2-M_m^2]},
\end{split}
\end{equation}
where
\begin{equation}
M_n = \sqrt{\Delta^2 + 2 n \mathcal{B}}
\end{equation}
are the energies of the relativistic Landau levels.
Deriving  Eqs.~(\ref{sigma-Pi12}) and (\ref{sigma-Pi34}) we used the fact that the
real part of $\Pi^{1,2,3,4}_{n,m}$ does not alter when the simultaneous replacements $i \Gamma \to - i \Gamma$ and  $i
\Gamma^\prime \to - i \Gamma^\prime$ are made, while its imaginary
part reverses sign.
Note that we used the notations  $\sigma_{ij}^{\textrm{H}}$ and $\sigma_{ij}^{\textrm{AH}}$
for the conductivities, although for its diagonal part $\sigma_{ij}^{\textrm{H}}$
simply corresponds to the standard part of the diagonal conductivity denoted in the main text as
$\sigma_{xx}^{\textrm{N}}$.

The Kronnecker delta symbols in Eqs.~(\ref{sigma-Pi12}) and (\ref{sigma-Pi34})
reflect the fact that the transitions only between the neighboring levels are involved.
Then we easily find the sum  over $m$ and obtain
\begin{widetext}
\begin{equation}
\begin{split}
\sigma_{ij}^{\textrm{H}}(\Omega)&=\frac{e^2 \mathcal{B} }{4\pi^2 \,\Omega} \sum\limits_{n=0}^\infty
\int\limits_{-\infty}^\infty
d\epsilon\left\{\delta_{ij}[n_F(\epsilon)-n_F(\epsilon^\prime)]{\rm Re}
[\Pi^1_{n,n+1}(\epsilon,\epsilon^\prime)+\Pi^1_{n+1,n}(\epsilon,\epsilon^\prime)-
\Pi^2_{n,n+1}(\epsilon,\epsilon^\prime)\right.\\
&-\left.\Pi^2_{n+1,n}(\epsilon,\epsilon^\prime)] -\epsilon_{ij}\,{\rm
sgn}(eB)\left[[n_F(\epsilon)-n_F(\epsilon^\prime)]{\rm Im}
[\Pi^1_{n,n+1}(\epsilon,\epsilon^\prime)-\Pi^1_{n+1,n}(\epsilon,\epsilon^\prime)]\right.\right.\\
&-\left.\left.[n_F(\epsilon)+n_F(\epsilon^\prime)]{\rm
Im}[\Pi^2_{n,n+1}(\epsilon,\epsilon^\prime)-\Pi^2_{n+1,n}
(\epsilon,\epsilon^\prime)]\right]\right\}
\end{split}
\end{equation}
and
\begin{equation}
\label{sigma-anomalous-general}
\begin{split}
\sigma_{ij}^{\textrm{AH}}(\Omega)&=\frac{e^2 \mathcal{B} }{4\pi^2 \,\Omega} \sum\limits_{n=0}^\infty
\int\limits_{-\infty}^\infty
d\epsilon\left\{\delta_{ij} {\rm sgn}(eB) [n_F(\epsilon)-n_F(\epsilon^\prime)]{\rm Re}
[\Pi^3_{n,n+1}(\epsilon,\epsilon^\prime)-\Pi^3_{n+1,n}(\epsilon,\epsilon^\prime)\right.\\
& \left. - \Pi^4_{n,n+1}(\epsilon,\epsilon^\prime) +\Pi^4_{n+1,n}(\epsilon,\epsilon^\prime)]
-\epsilon_{ij}\left[[n_F(\epsilon)-n_F(\epsilon^\prime)]{\rm Im}
[\Pi^3_{n,n+1}(\epsilon,\epsilon^\prime)+\Pi^3_{n+1,n}(\epsilon,\epsilon^\prime)]\right.\right.\\
&-\left.\left.[n_F(\epsilon)+n_F(\epsilon^\prime)]{\rm
Im}[\Pi^4_{n,n+1}(\epsilon,\epsilon^\prime)+\Pi^4_{n+1,n}
(\epsilon,\epsilon^\prime)]\right]\right\}.
\end{split}
\end{equation}
\end{widetext}
The remaining summation over $n$ can be performed expanding $\Pi^{1,2,3,4}$ in terms of the partial fractions.
Taking the dc limit, $\sigma_{ij} = \sigma_{ij}(\Omega \to 0)$, one obtains that the diagonal conductivity
is given by Eq.~(\ref{diagonal-conductivity}), where each term can be written in the form
(\ref{Kubo-diag}) with the corresponding kernel $\mathcal{D}_{\textrm{N,AH}}$.
\begin{widetext}
For the first term $\sigma_{xx}^{\textrm{N}}$, the kernel
\begin{equation}
\label{D-N}
\begin{split}
\mathcal{D}_\textrm{N}(\epsilon,\mathcal{B},\Gamma,\Delta) = & \frac{1}{2 \pi^2}
\frac{\Gamma^2}{\mathcal{B}^2 + 4 \epsilon^2 \Gamma^2}
 \left[ 2 \epsilon^2 + \frac{(\epsilon^2 + \Delta^2 +
\Gamma^2)\mathcal{B}^2 - 2 \epsilon^2 (\epsilon^2 - \Delta^2 +
\Gamma^2) \mathcal{B}}
{(\epsilon^2 - \Delta^2 - \Gamma^2)^2 + 4 \epsilon^2 \Gamma^2} \right. \\
&\left. - \frac{\epsilon(\epsilon^2 - \Delta^2 + \Gamma^2)}{\Gamma}
\mbox{Im} \psi \left( \frac{\Delta^2 - (\epsilon +  i \Gamma)^2}
{2 \mathcal{B}}\right) \right]
\end{split}
\end{equation}
was derived in Ref.~\onlinecite{Gorbar2002PRB}. For the anomalous part, $\sigma_{xx}^{\textrm{AH}}$,
we obtain
\begin{equation}
\label{sigma_xx-anomalous-kernel}
 \mathcal{D}_{\textrm{AH}}(\epsilon, \mathcal{B},\Gamma,\Delta) = \frac{\eta \Delta}{4 \pi^2}  \frac{ 2 \mathcal{B} \Gamma}
 {\mathcal{B}^2 + 4 \epsilon^2 \Gamma^2 }
 \left[\frac{2 \Gamma \epsilon (\mathcal{B}+ \Delta^2 + \Gamma^2 - \epsilon^2 ) }{[(\epsilon -\Delta)^2 + \Gamma^2] [(\epsilon +\Delta)^2 + \Gamma^2] }
+ \mbox{Im} \psi \left(\frac{\Delta^2 - (\epsilon + i \Gamma)^2}{2 \mathcal{B}} \right)
\right].
\end{equation}

The Hall conductivity $\sigma_{xy}(B,\Delta,\mu)$ can be obtained similarly to the diagonal conductivity.
The only difference is that after summation over $n$, there are terms that contain the Fermi distribution, $n_F(\epsilon)$.
They can be integrated by parts. The resulting Hall conductivity
is given by Eqs.~(\ref{Hall-general}) and (\ref{Kubo}) from the main text and the corresponding
anomalous Hall kernel is Eq.~(\ref{kernel-AH}). The normal Hall kernel
was derived in Ref.~\onlinecite{Gusynin2006PRB} and here we reproduce it for completeness \cite{foot2},
\begin{equation} \label{AH}
\begin{split}
& \mathcal{A}_{\textrm{H}}(\epsilon ,\mathcal{B},\Gamma,\Delta )= \frac{1}{4 \pi^2}
\left\{\frac{ \mathcal{B}(\Gamma^2-\Delta^2+\epsilon^2)}
{\mathcal{B}^2+4\epsilon^2\Gamma^2}\frac{2\epsilon\Gamma(\mathcal{B}+\Gamma^2+\Delta^2-\epsilon^2)}
{[\Gamma^2+(\Delta-\epsilon)^2][\Gamma^2+(\Delta+\epsilon)^2]}
+\frac{2\epsilon\Gamma}{\mathcal{B}}+\arctan\frac{\Delta+\epsilon}{\Gamma}
-\arctan\frac{\Delta-\epsilon}{\Gamma}\right.\\
& -\left.2{\rm
Im}\ln\Gamma\left(\frac{\Delta^2-(\epsilon+i\Gamma)^2}{2\mathcal{B}}\right)
+{\rm Im}\left[\left(\frac{\mathcal{B}(\Gamma^2-\Delta^2+\epsilon^2)}
{\mathcal{B}^2+4\epsilon^2\Gamma^2}+\frac{\Delta^2-(\epsilon+i\Gamma)^2}{\mathcal{B}}\right)
\psi\left(\frac{\Delta^2-(\epsilon+i\Gamma)^2}{2\mathcal{B}}\right)\right]\right\}.
\end{split}
\end{equation}
\end{widetext}
When the contribution of both $\eta = \pm $ valleys is summed, the anomalous Hall part (\ref{kernel-AH}) cancels out
and the remaining normal Hall kernel (\ref{AH}) leads to standard steplike dependence of the Hall
conductivity in the limit $\Gamma \to 0$ (see Appendix~B of Ref.~\onlinecite{Gusynin2006PRB}).

Finally, using the asymptotic of the digamma function,
\begin{equation}
\label{psi-asymp}
\psi(z)= \ln z  -
\frac{1}{2\,z} - \frac{1}{12\,z^2} + \frac{1}{120\,z^4}+
O\left(\frac{1}{z^5}\right)
\end{equation}
one obtains from the anomalous Hall kernel (\ref{kernel-AH})
its approximation (\ref{kernel-AH-appr}).

\section{Extraction of the oscillatory part}
\label{sec:Appendix-oscillations}

The kernels for the diagonal, $\mathcal{D}_{\textrm{N,AH}}$, and
Hall, $\mathcal{A}_{\textrm{H,AH}}$, conductivities have the oscillatory part contained
in the digamma function $\psi$ when the real part of its argument
becomes negative. Using the relationship
\begin{equation}
\psi(-z)=\psi(z)+ \frac{1}{z} + \pi \cot \pi z
\end{equation}
these oscillations in $1/B$ can be localized in the term with $\cot$.
Then we obtain for the imaginary and real parts of this term, respectively,
\begin{equation}
\label{osc-cot}
\begin{split}
&\binom{{\rm Im}}{{\rm Re}} \cot\left(\pi \frac{\epsilon^2-\Delta^2-\Gamma^2+2i\epsilon\Gamma}{2\mathcal{B}} \right)=\\
& \binom{- \sinh \left( \frac{2 \pi \epsilon \Gamma}{\mathcal{B}} \right) }{\sin \left(\pi \frac{\epsilon^2 - \Delta^2 - \Gamma^2}
{\mathcal{B}} \right)  }
\frac{1}{\cosh \left( \frac{2 \pi \epsilon \Gamma}{\mathcal{B}}\right)-
\cos \left(\pi \frac{\epsilon^2 - \Delta^2 - \Gamma^2}{\mathcal{B}} \right) }.
\end{split}
\end{equation}
For $\epsilon^2 > \Delta^2 + \Gamma^2$, taking the real and imaginary parts of the relationship,
\begin{equation}
\begin{split}
& \frac{e^{-(a-ib)}}{1-e^{-(a-ib)}}=\frac12\frac{\cos b-e^{-a}+i\sin b}{\cosh a-\cos b}\\
& =\sum_{n=1}^\infty e^{-n(a-ib)}
\end{split}
\end{equation}
one can expand, respectively, the top and bottom lines of Eq.~(\ref{osc-cot}).
In particular, one obtains the following expansion:
\begin{equation}\label{series}
\begin{split}
& \frac{\sinh (2 \pi |\epsilon| \Gamma/\mathcal{B})}{\cosh (2 \pi \epsilon
\Gamma/\mathcal{B}) - \cos [\pi(\epsilon^2 - \Delta^2 - \Gamma^2)/\mathcal{B}]} \\
& = 1 +
2 \sum_{n=1}^{\infty} \cos \left[ \frac{\pi n(\epsilon^2 - \Delta^2 -
\Gamma^2)}{\mathcal{B}} \right] \exp \left(- \frac{2 \pi n |\epsilon| \Gamma}{\mathcal{B}} \right).
\end{split}
\end{equation}
Using Eq.~(\ref{series}) one obtains from the kernel (\ref{kernel-AH}) the oscillatory part of the anomalous Hall term
\begin{widetext}
\begin{equation}
\label{AH-osc}
 \mathcal{A}_{\textrm{AH}}^{osc}(\omega,\mathcal{B},\Gamma,\Delta) = -\frac{\eta \Delta}{4 \pi^2}
\frac{8 \pi \Gamma^2 |\epsilon| }{\mathcal{B}^2 + (2 \epsilon \Gamma)^2}  \theta(\epsilon^2 - \Delta^2 - \Gamma^2)
  \sum_{n=1}^{\infty} \cos \left[ \frac{\pi n(\epsilon^2 - \Delta^2 -
\Gamma^2)}{\mathcal{B}} \right] \exp \left(- \frac{2 \pi n |\epsilon| \Gamma}{ \mathcal{B}} \right).
\end{equation}
In a similar fashion one can extract the oscillatory part from the last term of the normal Hall kernel
(\ref{AH}):
\begin{equation}
\label{H-osc}
\begin{split}
\mathcal{A}_{\textrm{H}}^{osc}(\epsilon ,\mathcal{B},\Gamma,\Delta )= &
- \frac{\mbox{sgn} \epsilon \, \theta(\epsilon^2 - \Delta^2 - \Gamma^2)}{2 \pi} \sum_{n=1}^\infty
\left\{\left[ \frac{\mathcal{B}(\Gamma^2-\Delta^2+\epsilon^2)}
{\mathcal{B}^2+4\epsilon^2\Gamma^2}+\frac{\Delta^2-\epsilon^2+ \Gamma^2}{\mathcal{B}}\right]
\cos \left[\frac{\pi n(\epsilon^2-\Delta^2-\Gamma^2)}{\mathcal{B}} \right]
\right.\\
& \left. + \frac{2 \epsilon \Gamma}{\mathcal{B}}
\sin \left[ \frac{ \pi n(\epsilon^2-\Delta^2-\Gamma^2)}{\mathcal{B}} \right] \right\}
\exp \left(-\frac{2\pi n|\epsilon|\Gamma}{ \mathcal{B}} \right).
\end{split}
\end{equation}
\end{widetext}
This shows that for a finite $\Delta$ in Eq.~(\ref{Hall-normal-osc}), the function
$g = g(\omega_c, \mu, \tau, \Delta)$.

The temperature dependence of the amplitude of the oscillations is acquired after the
corresponding kernel $\mathcal{D}_{\textrm{N,AH}}$ or
Hall, $\mathcal{A}_{\textrm{H,AH}}$ is integrated with the derivative of the
Fermi distribution, $-\partial n_F (\epsilon)/\partial \epsilon = (1/4T) \cosh^{-2} [(\epsilon -\mu)/2T]$.
Making a shift $\epsilon \to \epsilon + \mu$ and changing the variable $\epsilon \to 2 T \epsilon$, and
keeping only the linear in $T$ terms in the oscillating part of
the integrand, we use the integral
\begin{equation}
\label{integral}
\int_{0}^{\infty} d x \frac{\cos b x}{\cosh^2 x} = \frac{\pi
b/2}{\sinh \pi b/2}
\end{equation}
to obtain the temperature amplitude factor (\ref{temp-factor}).
In particular, one obtains from Eq.~(\ref{AH-osc}) the oscillatory part of
Eq.~(\ref{AH-oscil}). For $\Delta =0$ and small $\Gamma$, Eq.~(\ref{H-osc})
can be further simplified and one may present the final result for the Hall conductivity in the form of
Eq.~(\ref{Hall-normal-osc}), viz.,
\begin{equation}
\label{normal-Hall-Dirac}
\sigma^{\textrm{H}}_{xy}=-\frac{\sigma_0 \omega_c \tau \, \mbox{sgn} \, (eB) \, \mbox{sgn} \, \mu}{1+\omega_c^2 \tau^2}
\left[1-\frac{|\mu| \tau}{\omega_c^2 \tau^2}\frac{\tilde D(\mu)}{D_0(\mu)} \right],
\end{equation}
so that the function $g(\mu \tau)=|\mu|/(2\Gamma) = |\mu| \tau$.


\begin{thebibliography}{99}
\bibitem{foot2} Equation~(\ref{AH}) agrees with the corresponding
Eq.~(3.15) from Ref.~\onlinecite{Gusynin2006PRB}
if one takes into account that the  sum over the valley index $\eta$ is included in the latter.
\end{thebibliography}
\end{document}